\documentclass[a4paper]{article}

\usepackage{icrc2013}
\usepackage{color}
\usepackage[english]{babel}

\def\psim{\lower.5ex\hbox{$\; \buildrel \propto \over\sim \;$}}
\def\gtrsim{\lower.5ex\hbox{$\; \buildrel > \over\sim \;$}}
\def\lesssim{\lower.5ex\hbox{$\; \buildrel < \over\sim \;$}}

\email{charles.dermer@nrl.navy.mil, aws@mpe.mpg.de}
\shorttitle{ Spectrum of Cosmic Rays from Diffuse Gamma Rays}

\authors{
C.D.~Dermer$^{1}$,
A.W.~Strong$^{2}$,
E.~Orlando$^{3}$,
L.~Tibaldo$^{4}$,
for the $Fermi$-LAT  Collaboration.
}

\afiliations{
$^1$Code 7653, Naval Research Laboratory, 4555 Overlook Ave.\ SW, Washington, DC 20375 USA\\
$^2$Max-Planck-Institut f\"ur extraterrestrische Physik, Postfach 1312, D-85748 Garching, Germany\\
$^3$W.W. Hansen Experimental Physics Laboratory, Kavli Institute for Particle Astrophysics and Cosmology, Stanford University, Stanford, CA 94305, USA\\
$^4$Kavli Institute for Particle Astrophysics and Cosmology, SLAC National Accelerator Laboratory, 2575 Sand Hill Road, Menlo Park, CA 94025, USA\\
}

\begin{document}

\title{Determining the Spectrum of Cosmic Rays in  Interstellar Space from\\
 the Diffuse Galactic Gamma-Ray Emissivity}


%%\author{C.~D.\ Dermer, J.~D.\ Finke, R.~J.\ Murphy }
%\affiliation{Code 7653, Naval Research Laboratory, 4555 Overlook Ave.\ SW, Washington, DC 20375 USA}
%
%\author{A.~W.\ Strong}
%\affiliation{Max-Planck-Institut f\"ur extraterrestrische Physik, Postfach 1312, D-85748 Garching, Germany}
\author{F.\ Loparco, M.~N.\ Mazziotta}
%\affiliation{Istituto Nazionale di Fisica Nucleare, Sezione di Bari, 70126 Bari, Italy}
\author{E.\ Orlando}
%\affiliation{W.W. Hansen Experimental Physics Laboratory, Kavli Institute for Particle Astrophysics and Cosmology, Stanford University, Stanford, CA 94305, USA}
\author{T.\ Kamae,\footnote{Also at the Department of Physics, University of Tokyo, Tokyo, Japan} L.\ Tibaldo}
%\affiliation{Kavli Institute for Particle Astrophysics and Cosmology, Stanford Linear Accelerator Center, 2575 Sand Hill Road, Menlo Park, CA 94025, USA  }
%\author{J. Cohen-Tanugi}
%\affiliation{Laboratoire Univers et Particules de Montpellier, Universit\'e Montpellier 2, CNRS/IN2P3, F-34095 Montpellier, France }
\author{M. Ackermann}
%\affiliation{Deutsches Elektronen Synchrotron (DESY), D-15738 Zeuthen, Germany}
\author{T.\ Mizuno}
%\affiliation{Hiroshima Astrophysical Science Center, Hiroshima University, Higashi-Hiroshima, Hiroshima 739-8526, Japan}
%\author{F.~W.\ Stecker}
%\affiliation{NASA Goddard Space Flight Center, Greenbelt, MD 20771, USA}
%%\author{on behalf of the Fermi-LAT Collaboration}

\abstract
{
More than 90\% of the Galactic gas-related  $\gamma$-ray emissivity above 1 GeV
is attributed to  the decay of neutral 
pions formed in collisions between cosmic rays and interstellar matter, with lepton-induced processes 
becoming increasingly important below 1 GeV. 
Given the  high-quality measurements of the $\gamma$-ray 
emissivity of local interstellar gas 
between $\sim 50$ MeV and $\sim 40$ GeV obtained with the Large Area Telescope on board the {\it Fermi} space observatory,
it is timely to re-investigate this topic in detail, including the hadronic production mechanisms. 
 The emissivity spectrum will allow the interstellar cosmic-ray spectrum to be determined reliably,
providing a reference for origin and propagation studies as well as input to solar modulation models.
A method for such an analysis  and illustrative results are presented.
}
\maketitle


\section{Introduction}

The major part of the `diffuse' $\gamma$-ray emission of the Milky Way is of interstellar origin,
with the contribution from a superposition of unresolved, low luminosity
point sources estimated at only the 5-10\% level in the Galactic plane,
and with inverse-Compton emission contributing even less than that.
%subdominant.  [Subdominant to which--the major part, or the IC emission? --CD]
Almost all of the gas-related interstellar emission is 
 the result of   hadronic cosmic rays 
colliding with the nuclei of 
interstellar gas (see e.g. \cite{ste71,der86a,smr04,smp07}), and 
cosmic-ray electrons and positrons making bremsstrahlung $\gamma$ rays. Besides making $\gamma$-rays a valuable 
tracer of the interstellar matter and radiation, the spectral energy distribution of the diffuse emission
encodes the interstellar cosmic-ray spectrum. 

Using data taken in the first six months of {\it Fermi} science operations,
\cite{abdo09} measured the spectral energy distribution  of gamma-ray emission associated with local neutral atomic hydrogen, HI, between 100 MeV and 10 GeV, finding that it is proportional to the HI column density and deriving an emissivity (emission rate per H atom) spectrum.
Many measurements of
the HI emissivity in the neighborhood of the solar system and the outer Galaxy
were obtained using {\it Fermi} data. More recently, \cite{cas12} presented a new
emissivity spectrum of local atomic hydrogen between
Galactic latitudes  $10^\circ<|b|< 70^\circ$ for energies
$\sim 50$ MeV to $\sim 40$ GeV with 
error bars mainly  $\leq 15$\%.

It is therefore an optimal time to use the emissivity to deduce 
the interstellar cosmic ray proton spectrum with improved accuracy.

A review of near-threshold hadronic $\gamma$-ray production has been given in \cite{der13}, and more details will appear in \cite{str13}. Here we examine the total $\gamma$-ray production cross section in proton-proton collisions, remarking on the accuracy of different models in the low-energy, $< 10$ GeV, range. Some calculations are made for different production cross sections, illustrating the method. 
See  \cite{cas13} for an update on the local emissivity and an alternative approach to interpretation.
For a recent analysis of PAMELA data including solar modulation modelling and relation to the present work, see \cite{loparco}.

\section{$\gamma$ ray production cross section in proton collisions }

%%%%%%%%%%%%%%%%%%%%%%%%%%%%%%%%%%%%%%
\begin{figure*}
\centering
\resizebox{0.45\hsize}{!}{\includegraphics{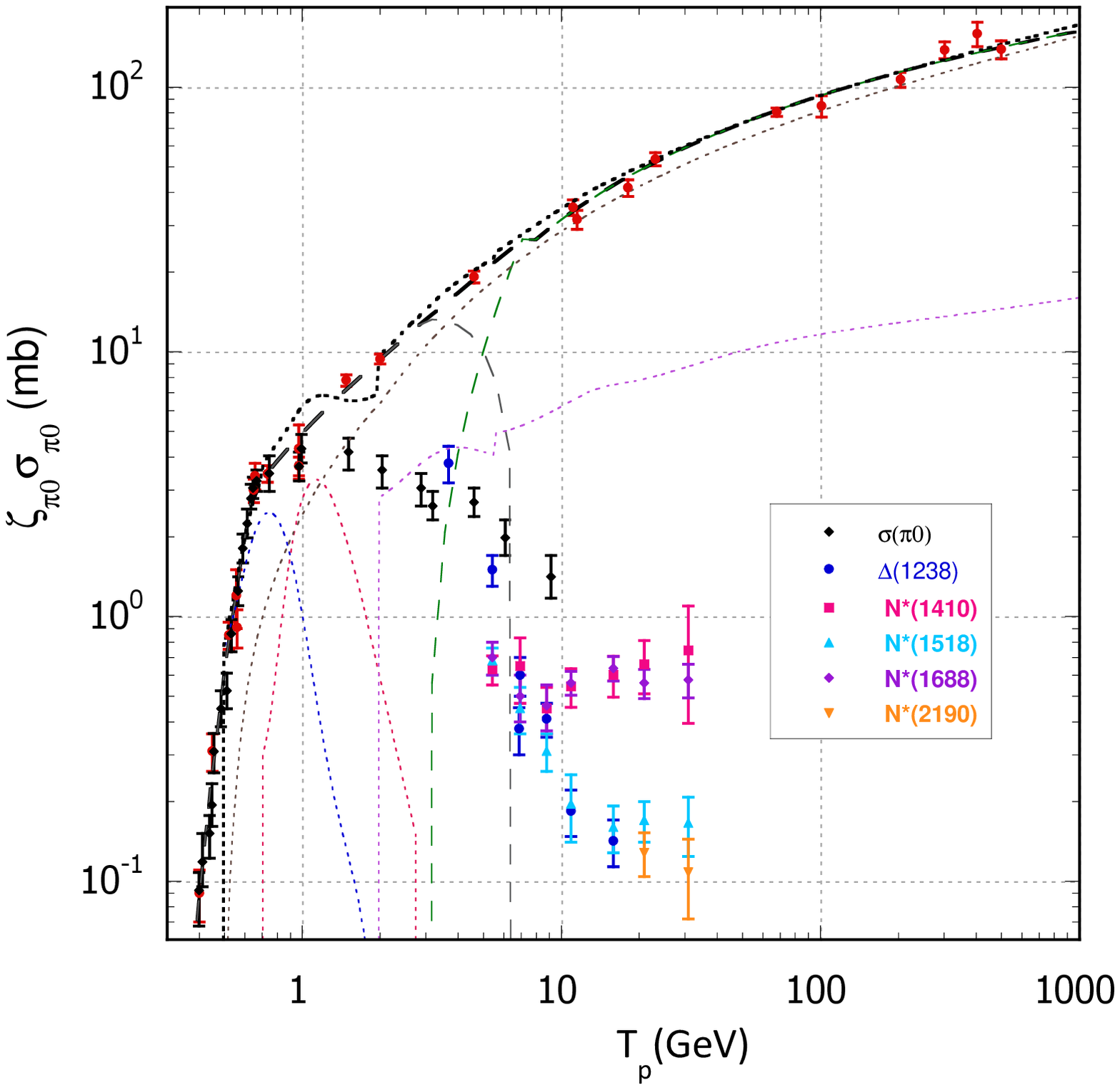}}
\resizebox{0.45\hsize}{!}{\includegraphics{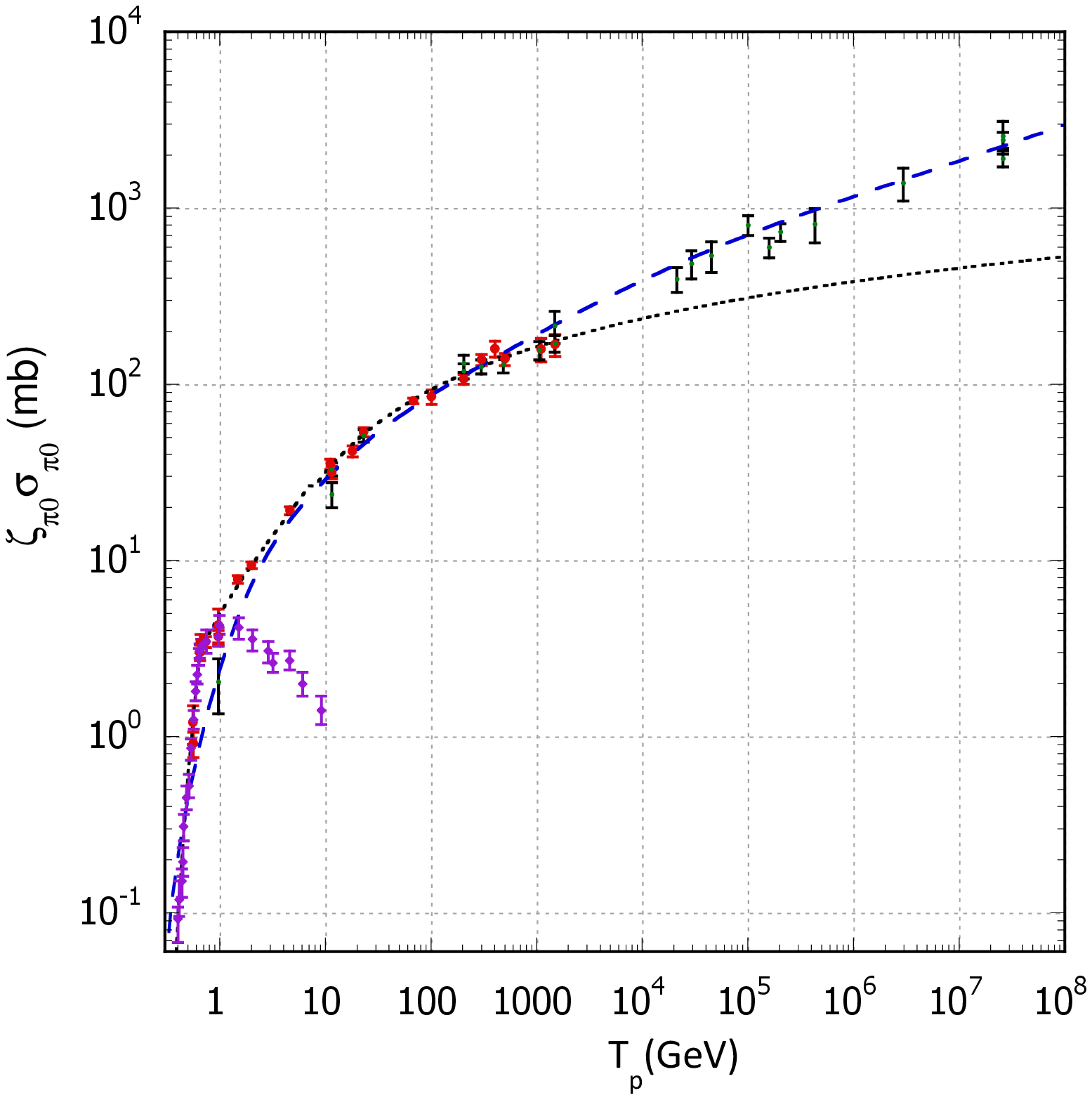}}
%{\bf \color{red} PRELIMINARY}
\caption{
Cross section data for inclusive $\pi^0$ and resonance production in p-p collisions.
({\it left}) Inclusive cross-section data (red points) are from \cite{ste73,der86b}, 
and cross-section data for single $\pi^0$ production, and $\Delta(1238)$, and $N^*$ resonances, as 
labeled, are taken from  \cite{lm70} and references therein. Component and total inclusive cross sections are shown by dashed and dotted curves for models of  \cite{der86a} and \cite{kam06}, respectively. ({\it right})  Inclusive cross-section data for pion production extending to high energies, including LHC data and fit from \cite{ssy12}, and fit of \cite{der86a}.
}
\label{results0}
\end{figure*}

The production threshold of $\pi^0$ ($m_\pi = 0.135$ GeV)
in collisions of energetic
protons with protons at rest is $2m_\pi + (m_\pi^2/2m_p)
=0.280$ GeV. Above this energy, the single $\pi^0$ cross section 
rises rapidly, reaching a value of $\approx 4$ mb above $T_p \approx 1.5$ GeV,
before declining to less than $\lesssim 1$~mb at $T_p \gtrsim 10$ GeV  \cite{lm70}.
Because of the rapidly declining cosmic-ray spectrum, most of the $\gamma$ 
rays are made from protons with $T_p\lesssim 5$ GeV \cite{der86a}, and a large fraction of these
$\gamma$ rays are made through resonance production. Almost all $\pi^0$ production below
$T_p\approx 1$ GeV is through the $\Delta(1238)$ isobar, and heavier isobars and resonances 
contribute at the per cent level to $T_p \gtrsim 30 $ GeV.

Fig.\ \ref{results0} shows data for inclusive $\pi^0$ production  \cite{ste73,der86b}, single $\pi^0$ production, and resonance
production   \cite{lm70} in $p~+~p$ collisions. Because of different isospin decay channels, only a fraction of the resonance cross section results in a $\pi^0$. 
The low- and high-energy dashed curves in the left panel of Fig.\ \ref{results0} represent, respectively, the $\Delta(1238)$ resonance and scaling contributions to the total inclusive $\pi^0$ cross section  in  the model of \cite{der86a}. The $\Delta(1238)$ resonance and $N^*(1600)$ resonance complex, and the diffractive and non-diffractive (scaling) contributions in the model of  \cite{kam06} are shown separately by the dotted curves. Model improvements, particularly at $T_p \lesssim 10$ GeV, are required to accurately infer the  interstellar cosmic-ray spectrum and its uncertainty.

%%%%%%%%%%%%%%%%%%%%%%%%%%
\section{Determining the interstellar cosmic-ray spectrum}
%%%%%%%%%%%%%%%%%%%%%%%%%%%%%%%%%%%%%%
\begin{figure*}
\centering
\resizebox{0.45\hsize}{!}{\includegraphics{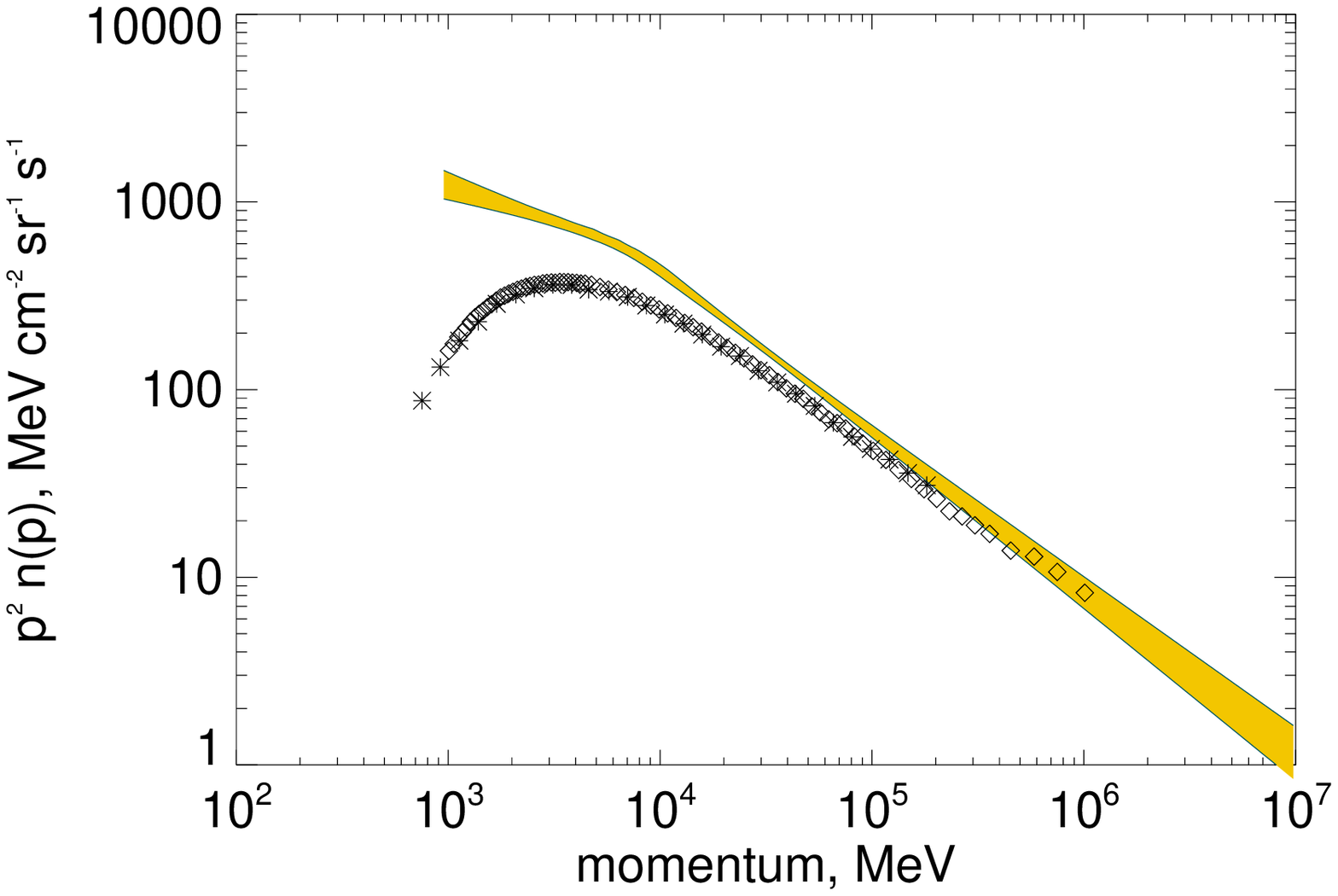}}
\resizebox{0.45\hsize}{!}{\includegraphics{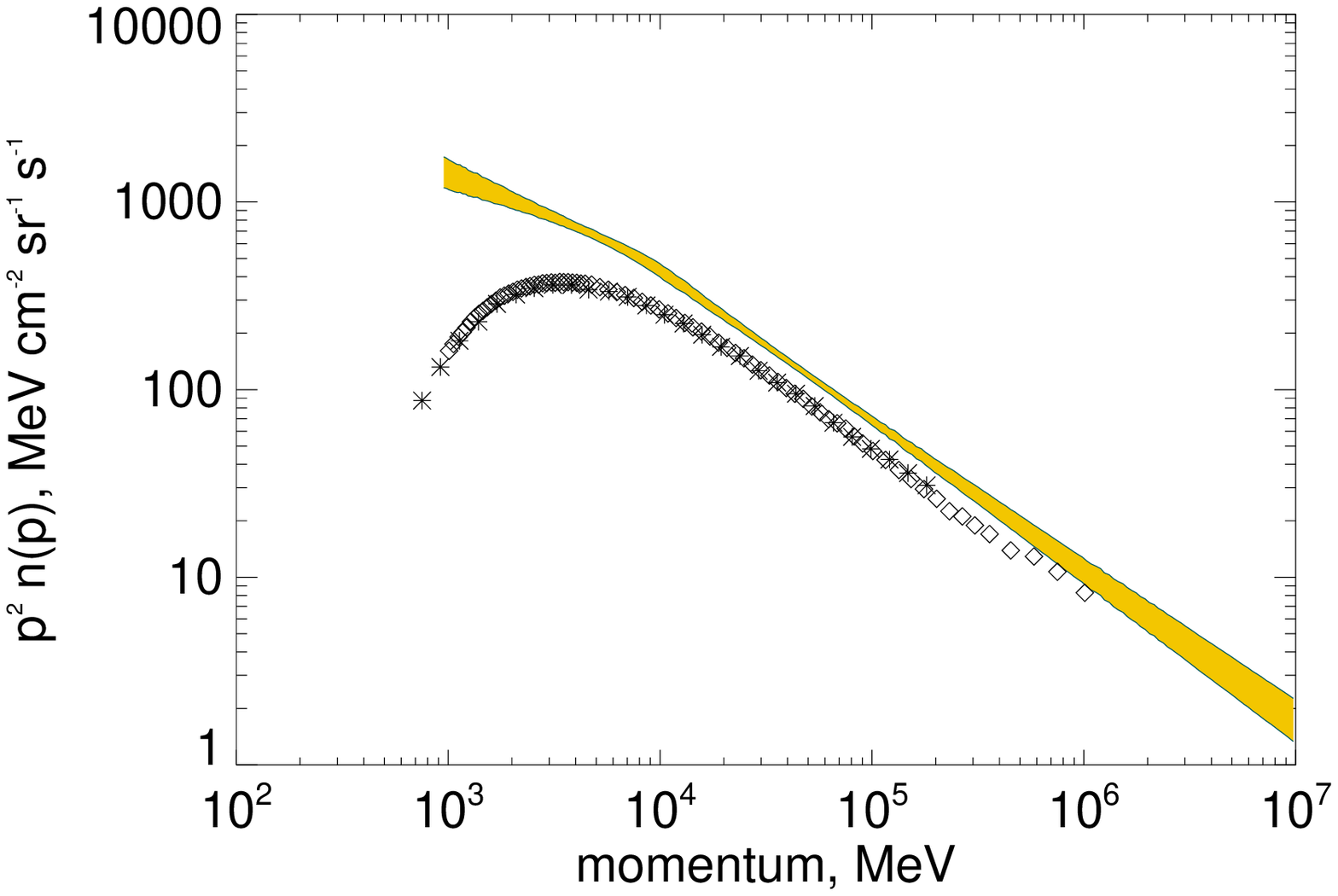}}
\resizebox{0.45\hsize}{!}{\includegraphics{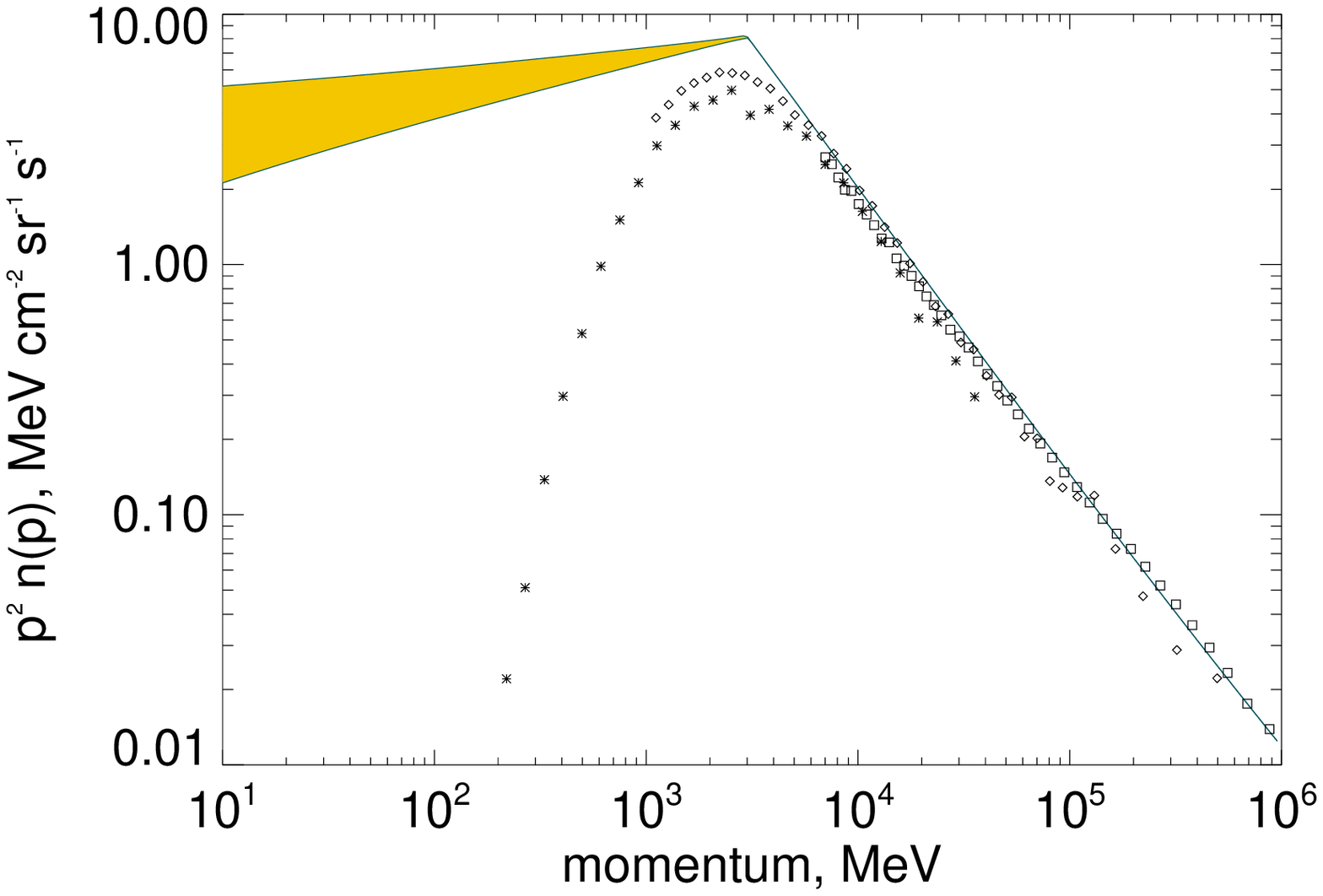}}
\resizebox{0.45\hsize}{!}{\includegraphics{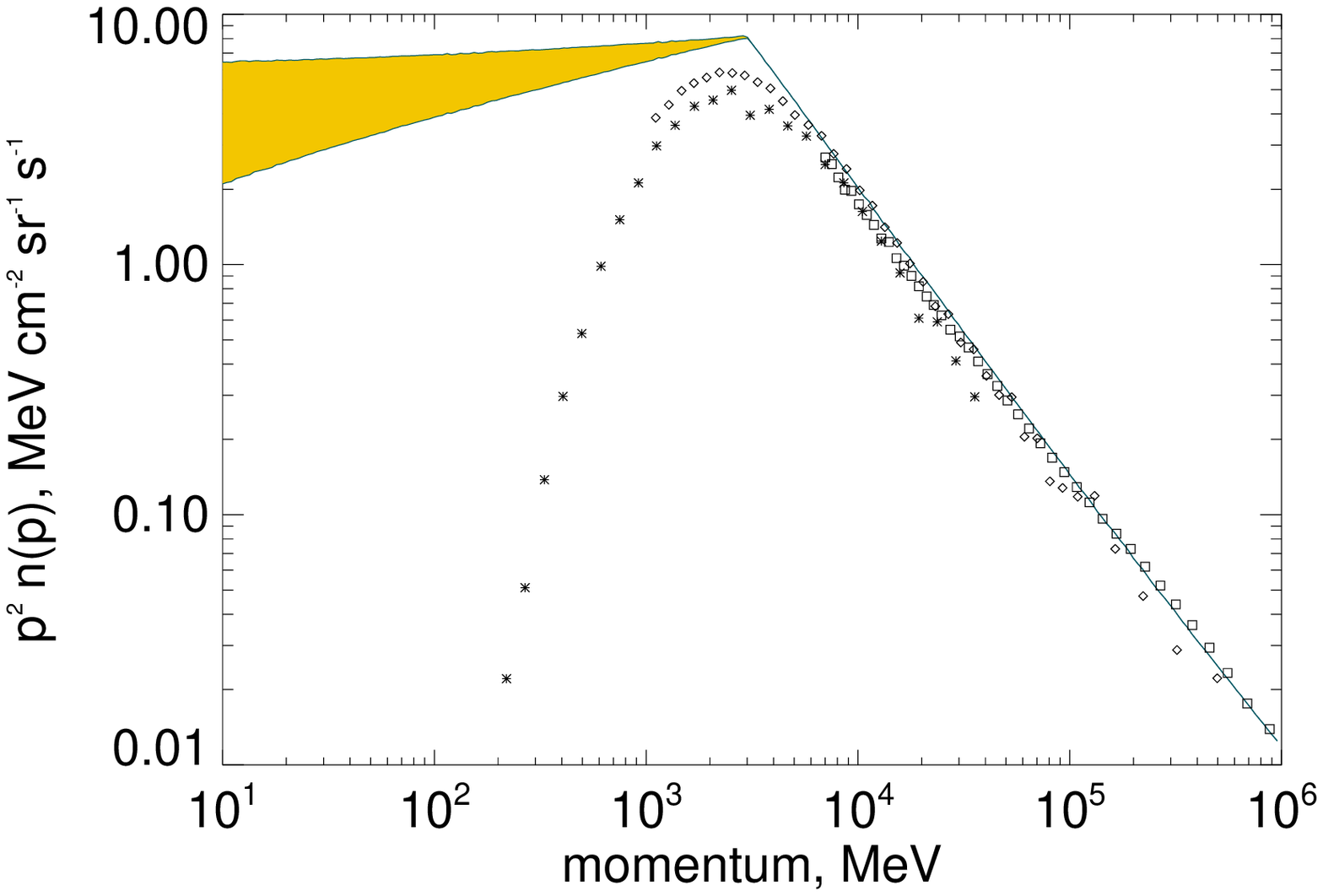}}
\resizebox{0.45\hsize}{!}{\includegraphics{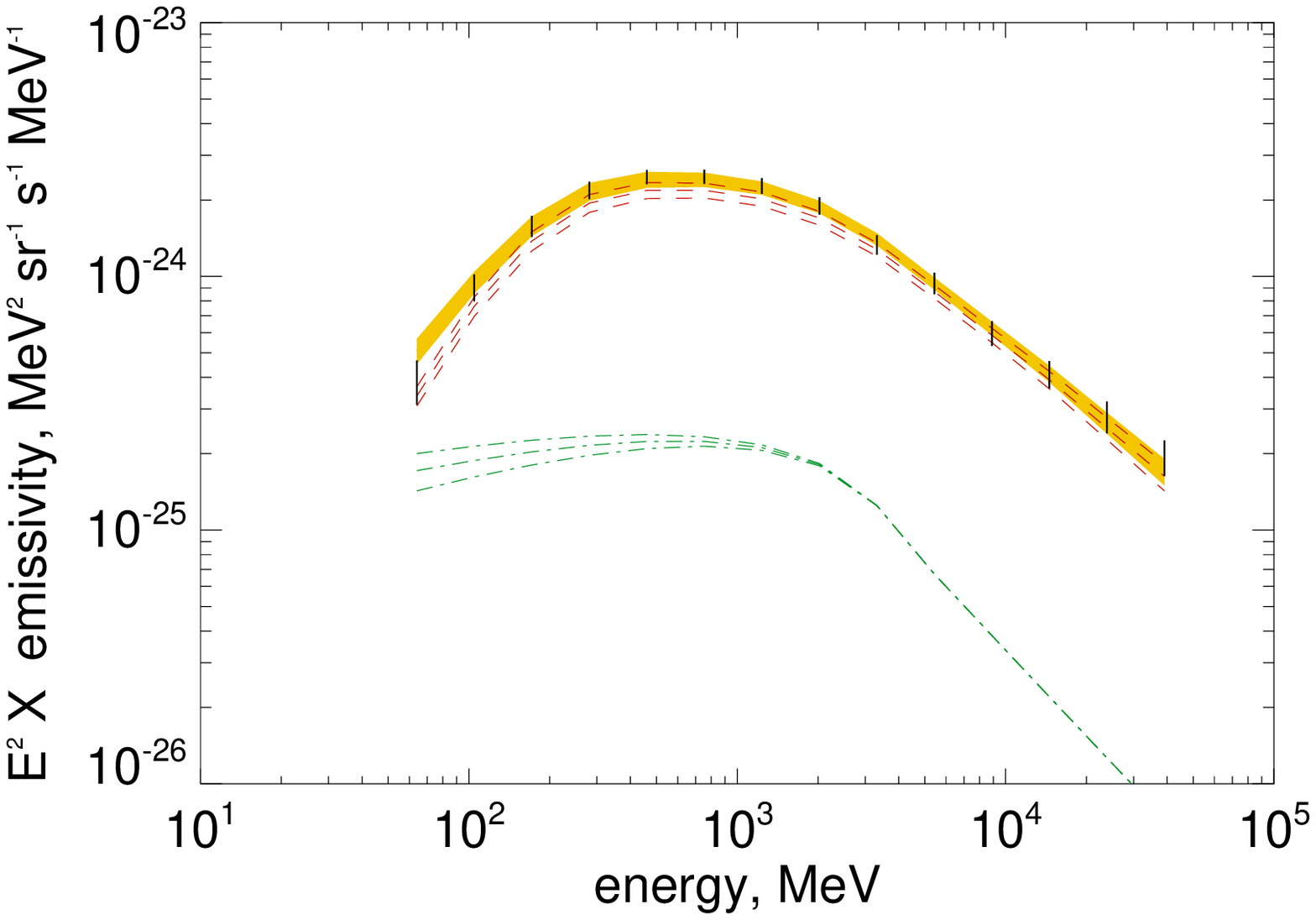}}
\resizebox{0.45\hsize}{!}{\includegraphics{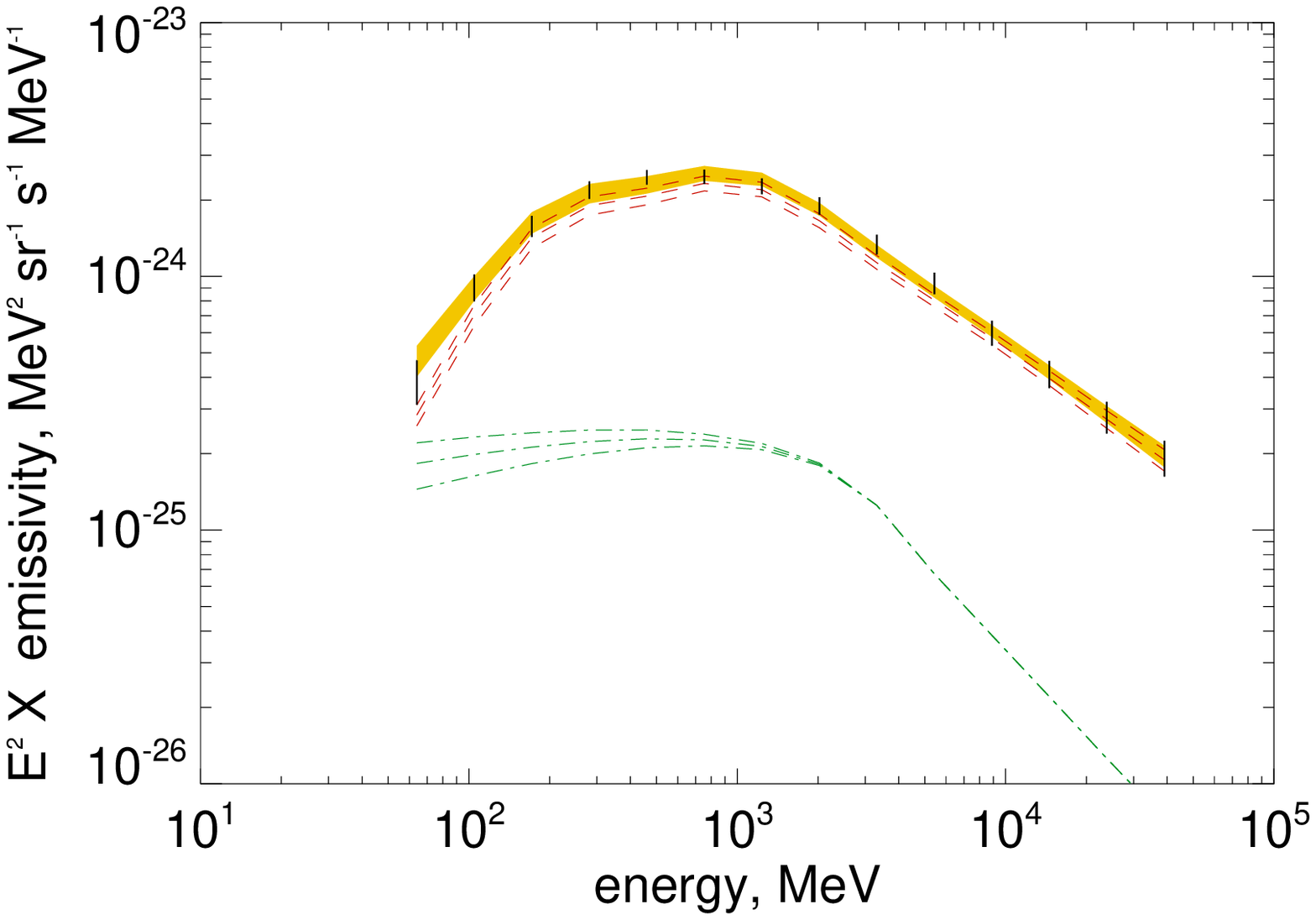}}
{\bf \color{red} PRELIMINARY}
\caption{Spectra derived from model fitting.
Yellow band shows model range. Model ranges are  1 standard deviation on the parameterized synthetic spectra.
Top: Measured and derived cosmic-ray proton spectra. Data are AMS01 (asterisks) and PAMELA (diamonds). 
Middle: Measured and derived cosmic-ray electron spectra. Data are AMS01 (asterisks), PAMELA (diamonds), and $Fermi$-LAT (squares). 
Bottom: $Fermi$-LAT emissivity data (vertical bars) and  model, with red and green curves showing the hadronic and leptonic  bremsstrahlung contributions; the yellow band shows the total. 
Emissivities from Casandjian \cite{cas12}, and cross sections from \cite{der86a} (left column), \cite{kam06,ko12} (right column). 
}
\label{results1}
\end{figure*}

Using the newly-measured emissivity spectrum \cite{cas12}, we can proceed to explore interstellar CR spectra that are compatible with it.
The analysis is performed by first computing the matrices connecting model CR spectra to the observed gamma-ray emissivities, in energy bands,
and then scanning the parameter space of the models.
The method is Bayesian, allowing a complete scan of the parameters, computing posterior probability distributions, mean values and error bars, while correctly accounting for the correlations among the parameters.

%%%%%%%%%%%%%%%%%%%%%%%%%%%%%%%%%%%%%%%%%%%%%%%%%%%%%

To illustrate the method, the measured emissivities have been fitted with $\gamma$-ray 
spectra calculated for a broken power law in momentum for protons and Helium,
with the free parameters being break momentum,  
spectral index below and above the break, and the overall normalization. 
The range of break momentum scanned is 1--20 GeV, since this is found necessary to fit the emissivity spectrum. 
The CR He/p ratio is fixed to the value measured by PAMELA at 100 GeV/nuc \cite{adr11},
and the He and p spectra are assumed to have the same shape (since they cannot be distinguished in gamma rays).
Note that CR spectra are expressed as particle density per momentum.\footnote{
Following  the physical motivation explained in \cite{der12}, we plot the density as a function of momentum $n(p)$, because 
%we multiply by $c/4\pi$ to be compatible with the usual experimental units at fully relativistic energies: 
the flux $j(T_p)=(\beta c/4\pi) n(T_p) = (\beta c/4\pi) |dp/dT_p| n(p) =  ( c/4\pi) n(p)$.}
The use of a sharp break in the CR spectrum is over-simplified but serves to illustrate the method; 
more physically plausible spectra (e.g., smooth breaks) are also being investigated,
but they do not lead to essentially different conclusions.

The first set  of hadronic cross sections used is from \cite{der86a}.  The p-p cross sections are scaled  for p-He, He-p and He-He interactions using the function given in \cite{norbury07}.
The second set  is from \cite{kam06} below 20 GeV, \cite{ko12} (QGSJET) above 20 GeV.
For  \cite{kam06} the p-p cross sections are scaled  for p-He, He-p and He-He interactions using \cite{norbury07}.
For  \cite{ko12} the  p-p, p-He, He-p cross-section are provided, so only  He-He is scaled from p-p.
The He fraction in the interstellar medium is taken as 0.1 by number.

The electron (plus positron) spectrum producing bremsstrahlung is based on $Fermi$-LAT measurements 
above 10 GeV \cite{2009PhRvL.102r1101A}, with a break below 3 GeV as indicated by synchrotron data \cite{soj11}.
The synchrotron data shown there require a flattening of the interstellar electron spectrum by about 
1 unit in the spectral index below a few GeV, so this is used as a constraint;
the actual low-energy index is determined by the  fit to the emissivities.

The resulting cosmic-ray proton and electron spectra and the corresponding emissivities are shown in Fig~\ref{results1}.
The fit to the measured emissivities is good, as can be expected with the freedom allowed.
The proton spectrum, having been determined from gamma rays (and gas tracers) alone with no input from direct CR measurements except for the He/p ratio, 
 is close to that measured directly at high energies. 
The solar modulation is clearly seen in the deviation of the interstellar spectrum from the direct measurements below 10 GeV.
Bremsstrahlung gives an essential contribution below $\approx 1$ GeV, and is an important component in  the analysis.

In this particular example, the interstellar proton spectrum 
steepens  by about 1/2 unit in the momentum index above a few GeV, compatible with expectations
from the cosmic-ray B/C ratio, which shows a similar break due to propagation. A power-law injection 
in momentum modified by propagation would then be a plausible scenario. 
The spectrum shown for \cite{der86a} cross sections has momentum index 2.5 (2.8) below (above) 6.5 GeV, with a scaling factor 1.4 relative to PAMELA at 100 GeV;
for  the \cite{kam06}, \cite{ko12} cross sections, the values are 2.4 (2.9) and 1.3, with the same break energy.
The formal significance of the break is about 4$\sigma$.
The sensitivity of the results to the cross sections is  significant but not overwhelming, 
though  uncertainties in the production cross sections must be included in the error budget
to derive firm conclusions.
In both of these illustrative cases, the high-energy proton index is compatible with PAMELA (2.82) \cite{adr11}.
The  scaling factor excess may have various origins, including uncertainties in the cross sections and the gas tracers, and/or hidden systematic errors in the direct measurements themselves. A difference between the interstellar spectrum and the direct measurements cannot be ruled out at this stage either.        

This preliminary result illustrates how it is possible to constrain the interstellar CR spectra with the $Fermi$-LAT emissivity data. % (cf.\ \cite{ner12}).
In a forthcoming paper \cite{str13}, all the uncertainties will be addressed, including those in emissivities (gas, instrumental response, etc.) 
and cross sections.  The evidence for a break and more exact constraints on the spectrum will be presented there.

\vskip0.1in
%%%%%%%%%%%%%%%%%%%%%%%%%%%%
The authors wish to thank Michael Kachelriess and Sergey Ostapchenko for valuable discussions on the use of  their cross-section code.
The work of C.D.D. supported by the Office of Naval Research and the NASA $Fermi$ Guest Investigator Program.

The $Fermi$ LAT Collaboration acknowledges support from a number of agencies and institutes for both development and the operation of the LAT 
as well as scientific data analysis. These include NASA and DOE in the United States, CEA/Irfu and IN2P3/CNRS in France, ASI and INFN in Italy, 
MEXT, KEK, and JAXA in Japan, and the K.~A.~Wallenberg Foundation, the Swedish Research Council and the National Space Board in Sweden. 
Additional support from INAF in Italy and CNES in France for science analysis during the operations phase is also gratefully acknowledged.

%\end{acknowledgments}

%\begin{thebibliography}{9}   % Use for  1-9  references
%%\begin{thebibliography}{99} % Use for 10-99 references

%\bibitem{accelconf-ref}
%http://www.cern.ch/accelconf

\end{document}